\documentclass[a4paper]{article}
\usepackage{amsmath}
\usepackage[english]{babel}
\usepackage{latexsym}
\usepackage{amssymb}
\usepackage{amscd,amsthm,amsfonts}
\usepackage{amsgen,amstext,amsbsy,amsopn}
\usepackage{epsfig}
\usepackage{subfigure}

\newcommand{\beq}{\begin{equation}}
 \newcommand{\eeq}{\end{equation}}
 \newcommand{\bdi}{\begin{displaymath}}
 \newcommand{\edi}{\end{displaymath}}
 \newcommand{\beqn}{\begin{eqnarray}}
 \newcommand{\eeqn}{\end{eqnarray}}
\newcommand{\al}{\alpha}
\newcommand{\g}{\gamma}
\newcommand{\f}{\frac}
\newcommand{\mc}{\mathcal}

\begin{document}


\title{Analytical galactic models with mild stellar cusps}
\author{\hspace{-.2 cm} T. Rindler-Daller\thanks{current address: Institut f\"ur Theoretische Physik, Universit\"at zu K\"oln,
  Z\"ulpicherstr.77, 50937 Cologne, Germany \newline email: trd@thp.uni-koeln.de}\\
\normalsize\it \hspace{-.5 cm}\\
\hspace{-.5 cm}\normalsize\it
    Fakult\"at f\"ur Physik, Universit\"at Wien, Boltzmanngasse 5, 1090 Vienna, Austria}

\date{to appear in \textit{Monthly Notices of the Royal Astronomical Society}}




\maketitle

\begin{abstract}
In the past two decades, it has been established by high-resolution
observations of early-type galaxies that their nuclear surface
brightness and corresponding stellar mass densities are
characterized by cusps. In this paper, we present a new spherical
analytical model family describing mild cuspy centres.
 We study isotropic and anisotropic models of Osipkov-Merritt
 type. It is shown that the associated distribution functions and intrinsic velocity dispersions can be
 represented analytically in a unified way in terms of
 hypergeometric series, allowing thus a straightforward comparison of these important global
quantities for galaxies having
 underlying mass densities which may differ significantly in their
 degree of central cuspiness or radial falloff.

\vspace{0,5cm}


\end{abstract}

\section{Introduction}

Since the early nineties of the last century, it has been
established by observations of ground- and space-based telescopes
that the nuclear surface brightness and corresponding stellar mass
densities of early-type galaxies are characterized by cusps. The
construction and study of galactic models incorporating cuspy
centres has therefore been an active part of theoretical modeling,
for which we can not give a full account here (see for instance
\cite{adams}, \cite{capuzzo}, \cite{cruz}, \cite{zeeuw},
\cite{leuwin}, \cite{sridhar}). An important issue for its own has
been the construction of analytical models describing cuspy
densities, even more so since up to the eighties most of the
analytical models available included (flat) cores only. One of the
first cuspy models, however, were given by \cite{jaffe} and
\cite{hern}. Further analytical spherical models with cusps have
been presented later e.g. by \cite{dehnen}, \cite{trem}, \cite{zhao}
and more recently by \cite{bd} and \cite{buyle}. However, many
models in the literature which are able to capture varying degrees
of cusps are rather inflexible with regard to the outer falloff
behaviour of the density or vice versa. Additionally, the associated
distribution functions must often be determined numerically, and
analytical expressions, if they exist, are often limited to a few
concrete values of the underlying model parameters. In this paper,
we like to present a very general spherical model family which
allows more flexibility in the central as well as in the outer
radial shape of the associated mass density, starting from a family
of non-singular, powerlaw-like potentials for the stellar component.
All further intrinsic quantities can be calculated analytically,
notably the distribution functions, for which an analytical
representation for a large range not mere for a little subset of
parameters is possible.
 This analyticity provides thus a straightforward study of the
relation between the cuspiness of the density and the behaviour of
the corresponding distribution function (and intrinsic velocity
dispersion), than it would be without analytical expressions at
hand. To this aim, we consider an isotropic as well as an
anisotropic parametrization of Osipkov-Merritt type for our models.
The projected quantities follow straightforwardly, but must be
determined numerically for our model family, unless the mass density
has a (flat) core. The family presented below is able to model the
cuspy centres of massive early-type galaxies and nucleated dwarf
elliptical galaxies. In addition, our family may also serve as
useful input for numerical studies on the time-dependent evolution
of galactic nuclei.
\\
This paper is organized as follows: In \textit{Section 2}, we
present our family of potentials and mass densities. In
\textit{Section 3}, we deduce the distribution functions for
isotropic and anisotropic Osipkov-Merritt parametrization. The
intrinsic velocity dispersions are calculated in \textit{Section 4}.
In \textit{Section 5}, we study the presence of a central
supermassive black hole in our model and in \textit{Section 6} we
present the conclusions. \textit{Appendix A} contains some often
used formulae.

\section{Model family}

We adopt the following family of spherical potentials
\begin{equation} \label{potpot}
\Phi(r) =
-\frac{\Phi_0}{\left(1+\left(\frac{r}{b}\right)^{\alpha}\right)^{\gamma}}
\end{equation}
with $\Phi_0$ and $b$ positive numbers and $\Phi(r) \sim -r^{-\al
\g}$ as $r \to \infty$. In fact, this family comprises almost all
non-singular powerlaw-like potentials found in the literature, most
of which are governed by one slope parameter. We choose the slope
parameters according to $0 < \alpha \leq 2$ and $\gamma
> 0$. In this paper, we
like to study the \textit{self-consistent} model originating from
this potential. From Poisson's equation follows the corresponding
mass density
\begin{displaymath}
\rho(r) =  \frac{\alpha \gamma b^{\alpha \gamma}\Phi_0}{4\pi G}
\frac{(1+\alpha)b^{\alpha}+(1-\alpha
\gamma)r^{\alpha}}{r^{2-\al}(b^{\alpha}+r^{\alpha})^{\gamma+2}}
\end{displaymath}
which is positive for $\alpha \gamma \leq 1$ and this restriction is
imposed throughout. The cuspiness of the density is determined by
the parameter $\alpha$. Flat cores with a central density of
$\rho(0) = 3\gamma \Phi_0/2\pi G b^2$ are recovered for $\al = 2$.
Otherwise, there is a cusp with $\rho(r) \sim r^{-2+\al}$. At large
radii, the density goes like $\rho(r) \sim r^{-2 -\al \g}$, hence
the degree of the outer falloff is governed by both parameters. The
above family includes a lot of known models as special cases: For
instance, the \cite{plum} model is recovered by setting $\al = 2, \g
= 1/2$. For $\al = 1$, the \cite{hern} model follows for $\g = 1$.
Other special cases obtained in the literature include $\al = 1/2,
\g = 2$ and $\al = 1, \g = \beta -3$ with $\beta \leq 4$, see
\cite{zhao}. In order to recover the cusps of the models of
\cite{dehnen} and \cite{trem}, one had to put $\al = 1/n, n \in
\mathbb{N}$. On the other hand, the outer falloff is recovered by
setting $\g = 1/n$. Both conditions at once can not be fulfilled to
recover the full models of the above authors. However, we note that
those models do not include mild cusps with $0 < 2-\al < 1$ whereas
our density does. \\
The associated cumulative mass function to
(\ref{potpot}) is given by
\begin{displaymath}
M(r) = \frac{\alpha \gamma \Phi_0 b^{\alpha
\gamma}}{G}\frac{r^{\alpha+1}}{(b^{\alpha}+r^{\alpha})^{\gamma+1}}
\end{displaymath}
 going for large radii as
  $M(r) \sim r^{1-\alpha \gamma}$, hence only models with $\alpha \gamma = 1$ have
   a finite total mass. In terms of the circular velocity
\begin{equation} \label{kreis}
v_{c}^{2} = \alpha \gamma b^{\alpha \gamma} \Phi_0
\frac{r^{\alpha}}{(b^{\alpha}+r^{\alpha})^{\gamma+1}},
\end{equation}
this amounts to $v_c \sim r^{-\al \g/2}$ for large $r$. Thus, the
circular velocity is Keplerian only for $\al \g = 1$, and decreases
more slowly for $\al \g < 1$. In the limit $\al \g \to 0$ it becomes
constant, $v_c \to const.$ However, for any fixed product $\al \g
\in (0,1)$, the increase in the cumulative mass is weaker as for the
logarithmic potential (\cite{binney}) or the isothermal sphere,
where $M(r) \sim r$ for large radii \footnote{On the other hand,
$M(r)$ diverges only logarithmically for the Hubble-Reynolds or
modified Hubble density profiles (see \cite{bt}).}. Self-consistent
models satisfying $\al \g < 1$ must therefore be cut off at some
outer radius in order to provide a finite mass. On the other hand,
due to its ability to reproduce constant or rising mass and velocity
profiles at large radii, the potential in (\ref{potpot}) may be
\textit{also} useful for modeling dark matter structures. In fact,
as an example we refer to \cite{wilk}, where the intrinsic
quantities for the stellar component were derived by assuming that a
dark matter component dominates the potential, the later of which is
a special case of (\ref{potpot}) with the parameters $\al \not= 0,
\g = 1$.
\\
 In the forthcoming, it is advantageous
to use dimensionless units: dividing (\ref{potpot}) by $-\Phi_0$ and
the mass density by $\al \g (1+\al)\Phi_0/4\pi G b^2$, we have for
the (relative) potential
\begin{equation} \label{inipot}
\Psi(r) = \frac{b^{\alpha \gamma}}{(b^{\alpha}+r^{\alpha})^{\gamma}}
\end{equation}
and for the density (using the same notation)
\begin{equation} \label{density}
\rho(r) = \frac{b^{\alpha \gamma +
2}}{1+\alpha}\frac{(1+\alpha)b^{\alpha}+(1-\alpha
\gamma)r^{\alpha}}{r^{2-\alpha}(b^{\alpha}+r^{\alpha})^{\gamma+2}}.
\end{equation}
These quantities will be used in subsequent calculations. We like to
put our emphasis on intrinsic quantities which can be calculated
analytically. Hence, the plots, which will be shown below, only
display the distribution functions and intrinsic velocity
dispersions, respectively. The corresponding projected quantities
for the family in (\ref{inipot}) \& (\ref{density})
 must be determined numerically except in the case of cores with $\alpha=2$, where analytical
 expressions in terms of hypergeometric functions can be given as well. In
 any case, it can be shown easily that the
surface brightness associated to (\ref{density}) rises steeply with
decreasing $\alpha$ as the projected radius tends to zero, since
then there is more stellar mass concentrated
 in the nuclear region. This behaviour is more pronounced for larger values of
 $\gamma$.

\section{Distribution functions}

As is shown in this and the following section, the distribution
functions (DFs) and intrinsic velocity dispersions for the above
model family (\ref{inipot}) \& (\ref{density}) can be calculated
analytically and we are going to study their behaviour for varying
cuspiness and outer falloff of the density (\ref{density}). To this
aim, we consider isotropic models with the DFs depending on the
relative energy $\mc{E} = -E$ as well as anisotropic Osipkov-Merritt
models, where they depend on $\mc{E}$ and the angular momentum $L$
via $Q = \mc{E} - L^2 /(2r_a^2)$ (see \cite{merritt} and
\cite{osipkov}).
 The anisotropy radius $r_a$ is a free parameter and the anisotropy function for this parametrization
 behaves as $\beta(r) = r^2/(r_a^2+r^2)$, hence the models are isotropic in the centres.\\
The isotropic DFs $f(\mc{E})$ for the model in (\ref{inipot}) \&
(\ref{density}) are calculated using Eddingtons's formula
\begin{equation} \label{eddi}
f(\mc{E}) =
\frac{1}{\sqrt{8}\pi^{2}}\frac{d}{d\mc{E}}\int_{0}^{\mc{E}}\frac{d\rho(\Psi)}{d\Psi}\frac{d\Psi}{\sqrt{\mc{E}
- \Psi}} =: \frac{1}{\sqrt{8}\pi^{2}}\frac{d}{d\mc{E}}I(\mc{E})
\end{equation}
by exploiting the fact that the density (\ref{density}) can be
expressed in terms of the potential (\ref{inipot}) as
\begin{displaymath}
\rho(\Psi) =
\frac{(\Psi^{-\f{1}{\gamma}}-1)^{1-\f{2}{\alpha}}}{1+\alpha}
\left[\alpha(1+\gamma)\Psi^{1+\f{2}{\gamma}} + (1-\alpha
\gamma)\Psi^{1+\f{1}{\gamma}}\right].
\end{displaymath}
The function $I(\mc{E})$ in (\ref{eddi}) is then
\begin{eqnarray} \label{integrale}
\lefteqn{I(\mc{E}) =
\frac{1}{1+\alpha}\left\{\alpha(1+\gamma)\left(1+\f{2}{\gamma}\right)\int_{0}^{\mc{E}}\f{\Psi^{\f{2}{\gamma}}
(\Psi^{-\f{1}{\gamma}}-1)^{1-\f{2}{\alpha}}}{\sqrt{\mc{E}-\Psi}}d\Psi
+ (1-\alpha
\gamma)\left(1+\f{1}{\gamma}\right)\int_{0}^{\mc{E}}\f{\Psi^{\f{1}{\gamma}}
(\Psi^{-\f{1}{\gamma}}-1)^{1-\f{2}{\alpha}}}
{\sqrt{\mc{E}-\Psi}}d\Psi \right. {} }
\nonumber\\
& &{} - \left. \frac{\alpha}{\gamma}
\left(1-\f{2}{\alpha}\right)(1+\gamma)\int_{0}^{\mc{E}}
\f{\Psi^{\f{1}{\gamma}}
(\Psi^{-\f{1}{\gamma}}-1)^{-\f{2}{\alpha}}}{\sqrt{\mc{E}-\Psi}}d\Psi
- \frac{1}{\gamma} \left(1-\f{2}{\alpha}\right)(1-\alpha
\gamma)\int_{0}^{\mc{E}}
\f{(\Psi^{-\f{1}{\gamma}}-1)^{-\f{2}{\alpha}}}
{\sqrt{\mc{E}-\Psi}}d\Psi \right\}.
\end{eqnarray}
The integrals in this expression can be determined analytically in
terms of Beta functions $B(a,b) = \Gamma(a) \Gamma(b)/\Gamma(a+b)$
and hypergeometric series $_q F_p (a_1,a_2,..,a_q;b_1,b_2,..,b_p;z)$
(see \cite{grad} for definitions and properties) \textit{provided}
that $2/\al = n, 1/\g = m$ and $2/(nm) \leq 1$ with $n,m \in
\mathbb{N}$. For core models having $\al = 2$, the only restriction
on the value of $\g$, however, is to be $\leq 1/2$. We use now
equation (\ref{ch47}) in the \textit{Appendix} to calculate the
integrals in (\ref{integrale}), which results into
 \beqn \label{intdf}
\lefteqn{I(\mc{E}) =
\frac{1}{1+\alpha}\left\{\mc{E}^{\f{1}{\g}\left(1+\f{2}{\al}\right)+\f{1}{2}}B\left(\f{1}{2},1+\f{2}{\al
\g}+\f{1}{\g}\right)
\left[\alpha(1+\gamma)\left(1+\f{2}{\gamma}\right)h_1\left(\f{2}{\al}-1\right)
- \frac{\alpha}{\gamma}
\left(1-\f{2}{\alpha}\right)(1+\gamma)h_1\left(\f{2}{\al}\right)\right]
\right. {} } \nonumber\\
& &{} + \left. \mc{E}^{\f{2}{\al
\g}+\f{1}{2}}B\left(\f{1}{2},\f{2}{\al \g}+1\right) \left[(1-\al
\gamma)\left(1+\f{1}{\gamma}\right)h_2\left(\f{2}{\al}-1\right) -
\frac{1}{\gamma} \left(1-\f{2}{\alpha}\right)(1-\al
\gamma)h_2\left(\f{2}{\al}\right)\right] \right\},
 \eeqn
where we defined for brevity
 \bdi
  h_1(x) := ~_{\f{1}{\g}+1}F_{\f{1}{\g}}\left(x,1+\f{2}{\al}+\g,1+\f{2}{\al}+2\g,...,
2+\f{2}{\al}; 1+\f{2}{\al}+\f{3\g}{2},1+\f{2}{\al}+\f{5\g}{2},
...,2+\f{2}{\al}+\f{\g}{2};\mc{E}^{\f{1}{\g}}\right)
 \edi
  and
 \bdi
  h_2(x) := ~_{\f{1}{\g}+1}F_{\f{1}{\g}}\left(x,\f{2}{\al}+\g,\f{2}{\al}+2\g,...,1+\f{2}{\al};
\f{2}{\al}+\f{3\g}{2},\f{2}{\al}+\f{5\g}{2},...,1+\f{2}{\al}+\f{\g}{2};\mc{E}^{\f{1}{\g}}\right).
 \edi
In order to calculate the derivative of (\ref{intdf}), we use the
general relation (\ref{hypergeom}). Abbreviating
  \bdi
 h_3(x) := ~_{\f{1}{\g}+1}F_{\f{1}{\g}}\left(x,2+\f{2}{\al}+\g,2+\f{2}{\al}+2\g,...,3+\f{2}{\al};2+
\f{2}{\al}+\f{3\g}{2},2+\f{2}{\al}+\f{5\g}{2},...,3+\f{2}{\al}+\f{\g}{2};\mc{E}^{\f{1}{\g}}\right),
  \edi
we finally arrive at the expression for the isotropic distribution
function
 \beqn \label{dist}
 \lefteqn{f(\mc{E}) =
(\sqrt{8}\pi^2(1+\al)\g)^{-1} \mc{E}^{-1/2} \times {}}
\nonumber\\
 & &{}
 \left\{ B\left(\f{1}{2},\f{1}{\g}+\f{2}{\al \g}+1\right) \al (1+\g) \left(\mc{E}^{\f{1}{\g}(1+\f{2}{\al})}
 \left(\f{\g}{2} + 1 + \f{2}{\al}\right)
 \left[\left(1+\f{2}{\g}\right)h_1\left(\f{2}{\al}-1\right)-
 \f{1}{\g}\left(1-\f{2}{\al}\right)h_1\left(\f{2}{\al}\right)\right] + \right. \right.\nonumber\\
 & &{}
 \left. \left. + \mc{E}^{\f{2}{\g}(1+\f{1}{\al})}
  \f{\left(1+\f{2}{\al}+\g\right)\left(1+\f{2}{\al}+2\g\right)...
  \left(2+\f{2}{\al}\right)}{\left(1+\f{2}{\al}+\f{3\g}{2}\right)\left(1+\f{2}{\al}+\f{5\g}{2}\right)
  ...\left(2+\f{2}{\al}+\f{\g}{2}\right)}
  \left[\left(1+\f{2}{\g}\right)\left(\f{2}{\al}-1\right)h_3\left(\f{2}{\al}\right)-
\f{1}{\g}\left(1-\f{2}{\al}\right)\f{2}{\al}h_3\left(\f{2}{\al}+1\right)\right]\right)
\right.
\nonumber\\
& &{}
 + \left.
B\left(\f{1}{2},\f{2}{\al \g}+1\right) (1-\al \g)
\left(\mc{E}^{\f{2}{\al \g}}\left(\f{\g}{2} + \f{2}{\al}\right)
 \left[\left(1+\f{1}{\g}\right)h_2\left(\f{2}{\al}-1\right)-
 \f{1}{\g}\left(1-\f{2}{\al}\right)h_2\left(\f{2}{\al}\right)\right] + \right. \right. \nonumber\\
 & &{}
 \left. \left. + \mc{E}^{\f{1}{\g}(1+\f{2}{\al})}
  \f{\left(\f{2}{\al}+\g\right)\left(\f{2}{\al}+2\g\right)...
  \left(1+\f{2}{\al}\right)}{\left(\f{2}{\al}+\f{3\g}{2}\right)\left(\f{2}{\al}+\f{5\g}{2}\right)
  ...\left(1+\f{2}{\al}+\f{\g}{2}\right)}
\left[\left(1+\f{1}{\g}\right)\left(\f{2}{\al}-1\right)h_1\left(\f{2}{\al}\right)-
\f{1}{\g}\left(1-\f{2}{\al}\right)\f{2}{\al}h_1\left(\f{2}{\al}+1\right)\right]\right)\right\}.
 \eeqn
This functions involves powers of $\mc{E}$ multiplied by
hypergeometric series, the later of which may even reduce to simpler
analytical functions of $\mc{E}$ depending on the values for $\al$
and $\g$. The order of the hypergeometric functions is determined by
$\gamma$. The DFs for models with cores, $\al = 2, \g \leq 1/2$,
simplify considerably and are given by
 \bdi
  f(\mc{E}) =
  \f{1}{3\sqrt{8}\pi^2}\left[2(1+\g)\left(1+\f{2}{\g}\right)\left(\f{1}{2}+\f{2}{\g}\right)B\left(\f{1}{2},\f{2}{\g}+1\right)
  \mc{E}^{\f{2}{\g}-\f{1}{2}} + (1-2\g)\left(1+\f{1}{\g}\right)\left(\f{1}{2}+\f{1}{\g}\right)B\left(\f{1}{2},\f{1}{\g}+1\right)
  \mc{E}^{\f{1}{\g}-\f{1}{2}}\right].
 \edi
The function in (\ref{dist}) is plotted in Fig. 1, first row, left
plot, for the models $\al = 2, \g = 1/3;~ \al = 1, \g = 1/2;~ \al =
2/3, \g = 1;~ \al = 1/2, \g = 1/2$. As a result of the finite depth
of the central potential well, $\Psi(0) = \mc{E}_{max} = 1$, the
distribution functions diverge for $\mc{E} \to 1$: As $\mc{E}
\rightarrow 1$, a steeper inner cusp corresponds to a stronger
 divergence in this limit because the system is then dominated by
 stars
 at small radii where the cusp dominates and this effect is therefore hardly affected by
 $\gamma$. On the other hand, the decrease of $f(\mc{E})$ as $\mc{E} \rightarrow 0$ is
 larger for small values of $\gamma$. This is more pronounced if $\alpha$ is small as
  well because then the model is more centrally concentrated as a result of the cusp.
\\
Now we turn to the anisotropic models: The Osipkov-Merritt
distribution functions for the model family in (\ref{inipot}) \&
(\ref{density}) can be calculated accordingly from a similar
relation as the one given in (\ref{eddi}), (see \cite{carollo}),
namely
\begin{displaymath}
f_a(Q) =
\frac{1}{\sqrt{8}\pi^{2}}\frac{d}{dQ}\int_{0}^{Q}\frac{d\rho_{a}}{d\Psi}\frac{d\Psi}{\sqrt{Q
- \Psi}}
\end{displaymath}
using the auxiliary density $\rho_{a}(r) := \left(1 +
\frac{r^{2}}{r_{a}^{2}}\right)\rho(r)$. The result is
 \bdi
f_a(Q) = f(Q) +
\f{1}{\sqrt{8}\pi^2}\f{1}{1+\al}\left(\f{b}{r_a}\right)^2 Q^{-1/2}
\times
 \edi
 \beq \label{adf}
\left\{[\al(2\g+1)-1]\left(1+\frac{1}{\g}\right)\left(\frac{1}{2}+\frac{1}{\g}\right)B\left(\frac{1}{2},1+\f{1}{\g}\right)
Q^{\f{1}{\g}}
-\al(1+\g)\left(1+\f{2}{\g}\right)\left(\f{1}{2}+\f{2}{\g}\right)B\left(\f{1}{2},1+\f{2}{\g}\right)
Q^{\f{2}{\g}} + 1-\al \g\right\}. \eeq
 The first term $f(Q)$ is given
by the expression for the isotropic DFs in (\ref{dist}) except that
$\mc{E}$ has to be replaced everywhere by $Q$. The anisotropic DF
(\ref{adf}) is plotted in Fig. 1, first row, right plot, using the
same model parameters as before. For the same reason as in the
isotropic case, the increase of $f_a(Q)$ for $Q \rightarrow 1$ is
dominated by the cusp parameter $\alpha$.
 On the other hand, for fixed $\alpha$ the parameter $\gamma$ controls essentially the degree
 of the anisotropy in the sense that the model is more anisotropic for small
  values of $\gamma$. As a general result we see that the anisotropic DFs do not decrease as
  rapidly for $Q \rightarrow 0$ as do the isotropic DFs: It can be easily shown that the
models
    approach the isotropic behaviour for large $r_a > b$, as expected. In contrast, for $r_{a} < b$
    the anisotropic signature in $f_a(Q)$ dominates over a wider range in $Q$, whereas
     the increase for $Q \rightarrow 1$ remains quite unaffected.
\begin{figure*}
\begin{minipage}[b]{0.4\linewidth}
      \centering\includegraphics[angle=270,width=5cm]{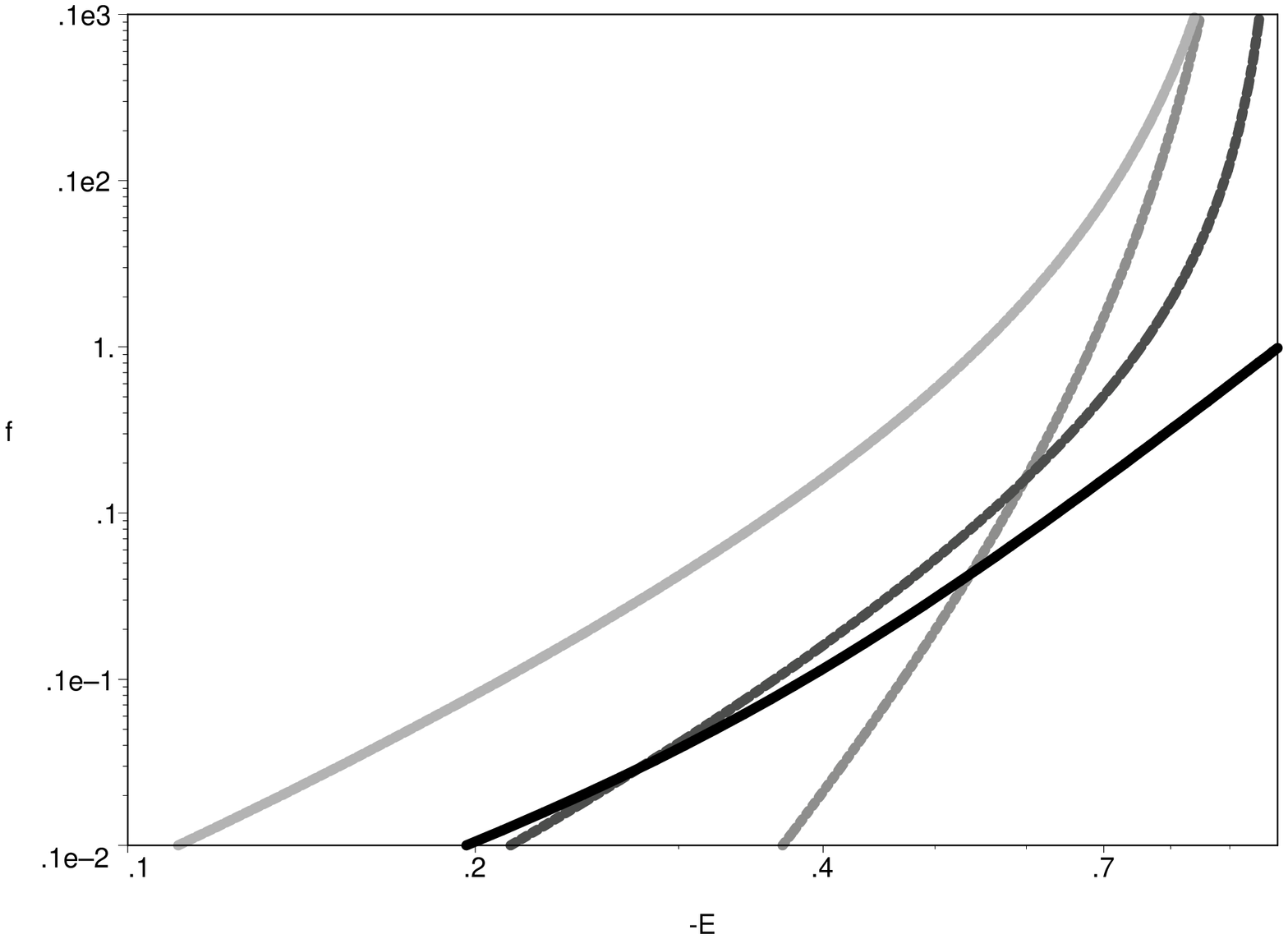}
     \hspace{0.3cm}
    \end{minipage}%
    \vspace{0.5cm}
    \begin{minipage}[b]{0.4\linewidth}
      \centering\includegraphics[angle=270,width=5cm]{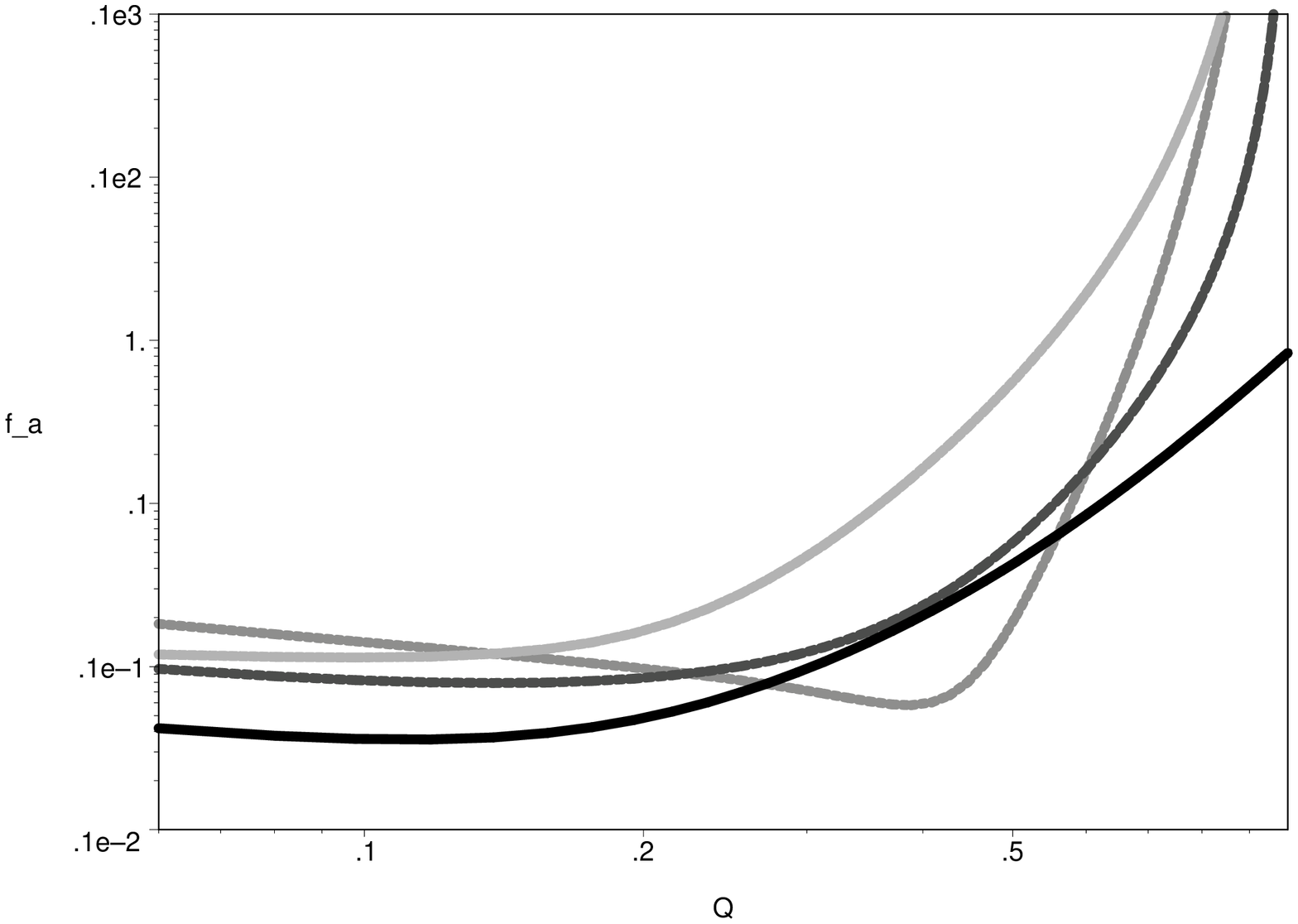}
     \hspace{0.3cm}
    \end{minipage}
    \vspace{0.5cm}
\begin{minipage}[b]{0.33\linewidth}
      \centering\includegraphics[angle=270,width=4.9cm]{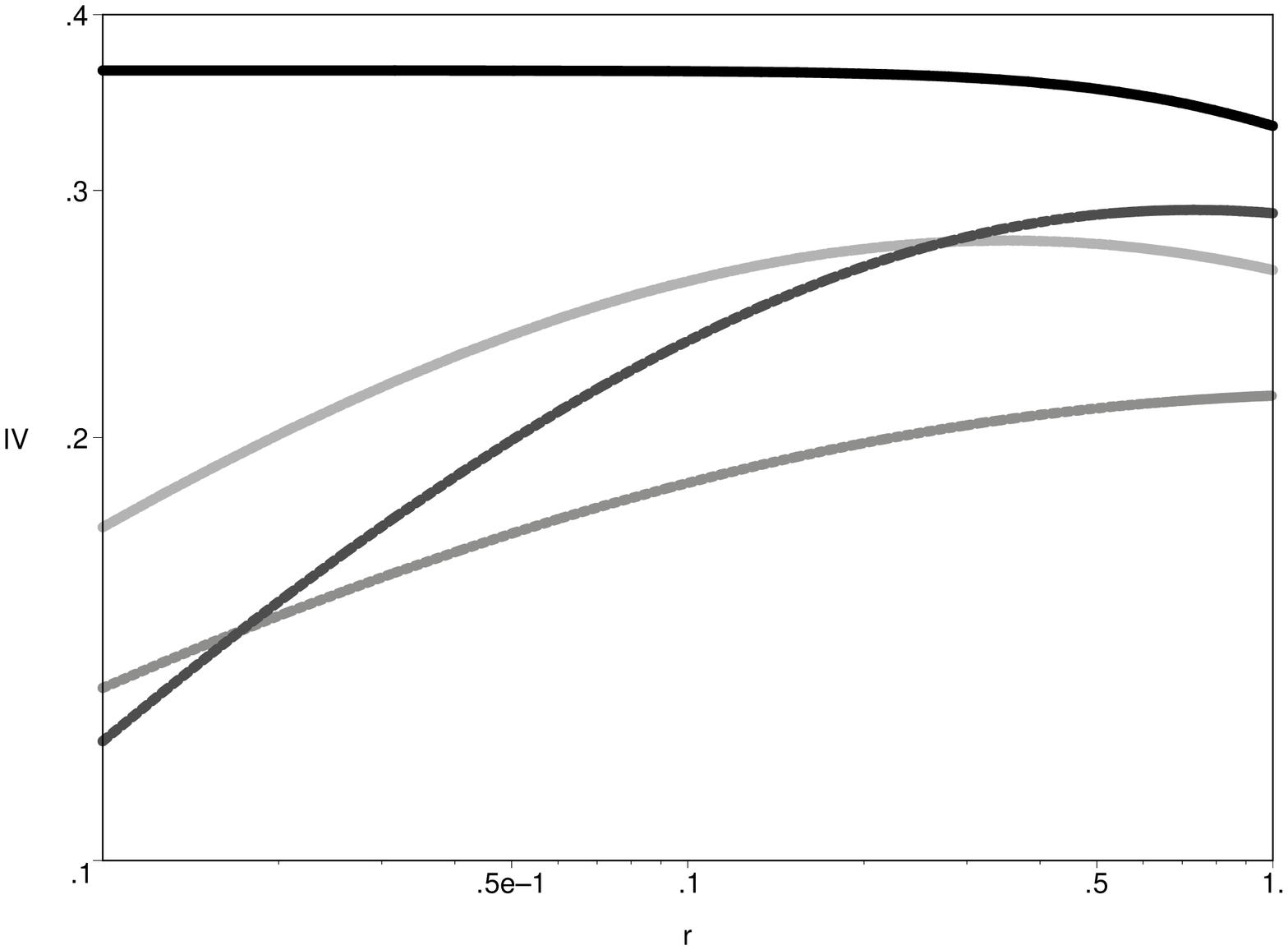}
     \hspace{0.3cm}
    \end{minipage}%
    \begin{minipage}[b]{0.33\linewidth}
      \centering\includegraphics[angle=270,width=5cm]{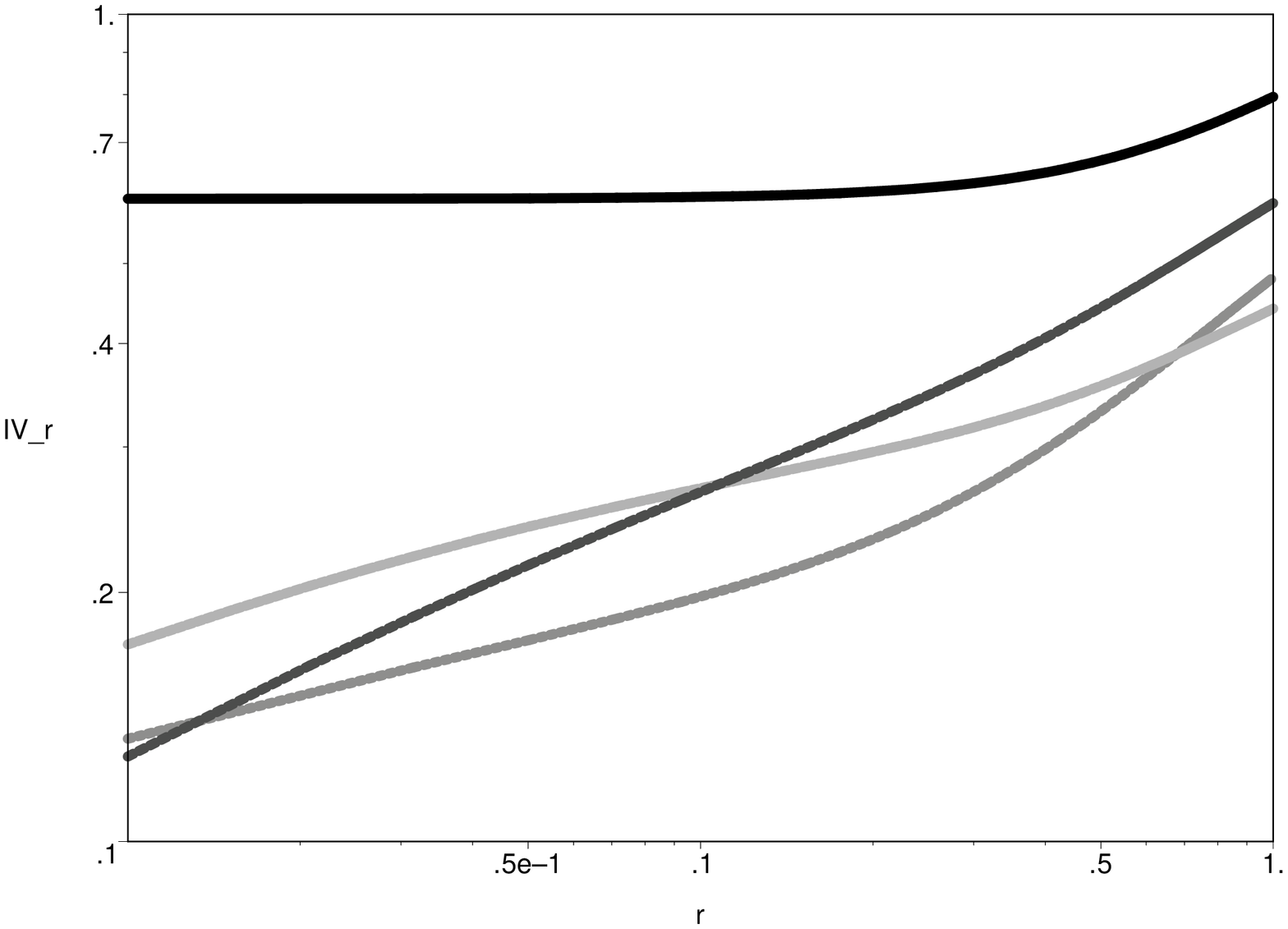}
     \hspace{0.3cm}
    \end{minipage}
    \begin{minipage}[b]{0.33\linewidth}
      \centering\includegraphics[angle=270,width=5cm]{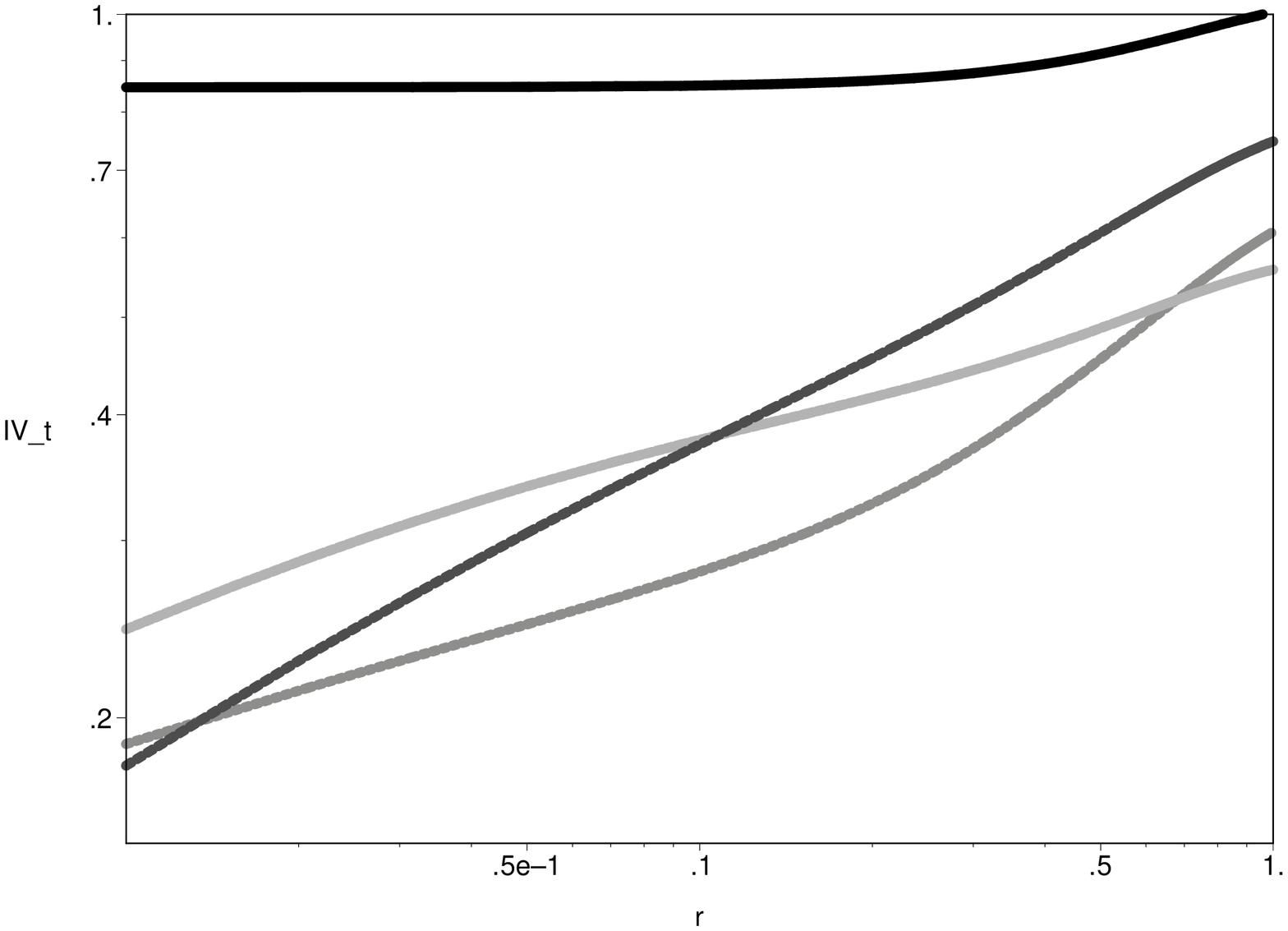}
     \hspace{0.3cm}
    \end{minipage}
\begin{minipage}[b]{0.33\linewidth}
      \centering\includegraphics[angle=270,width=5cm]{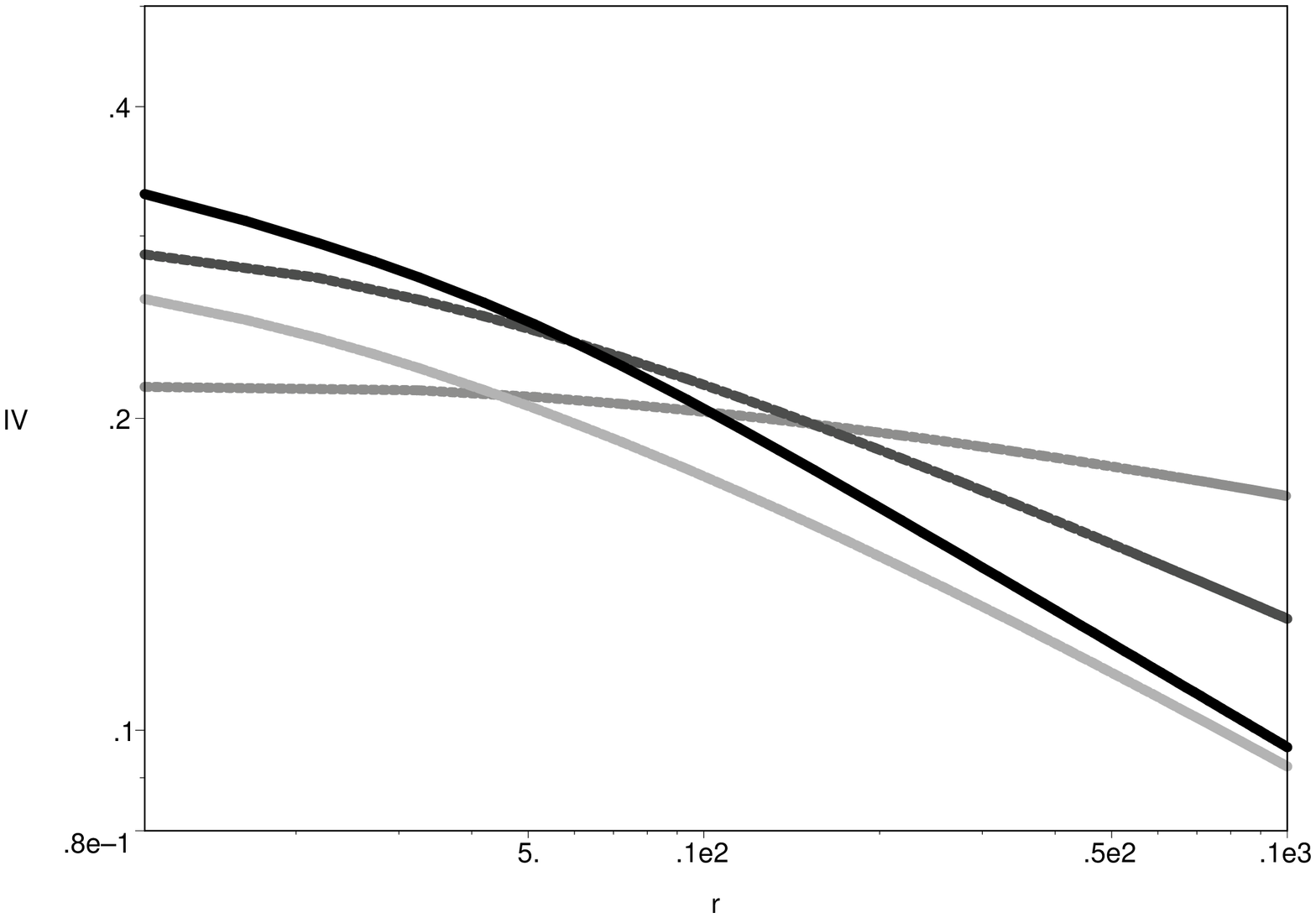}
     \hspace{0.3cm}
    \end{minipage}%
     \begin{minipage}[b]{0.33\linewidth}
      \centering\includegraphics[angle=270,width=5cm]{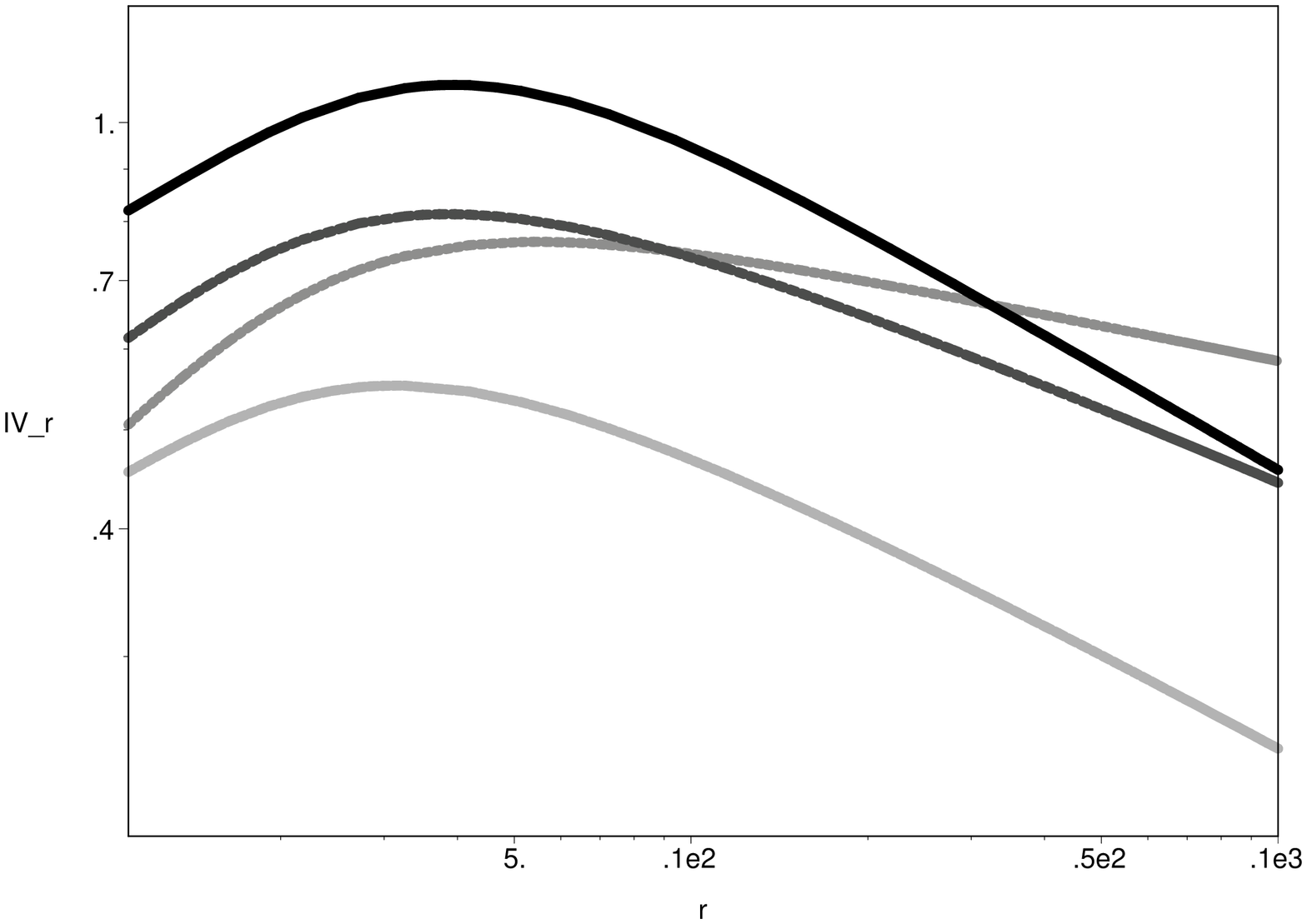}
     \hspace{0.3cm}
    \end{minipage}
    \begin{minipage}[b]{0.33\linewidth}
      \centering\includegraphics[angle=270,width=5cm]{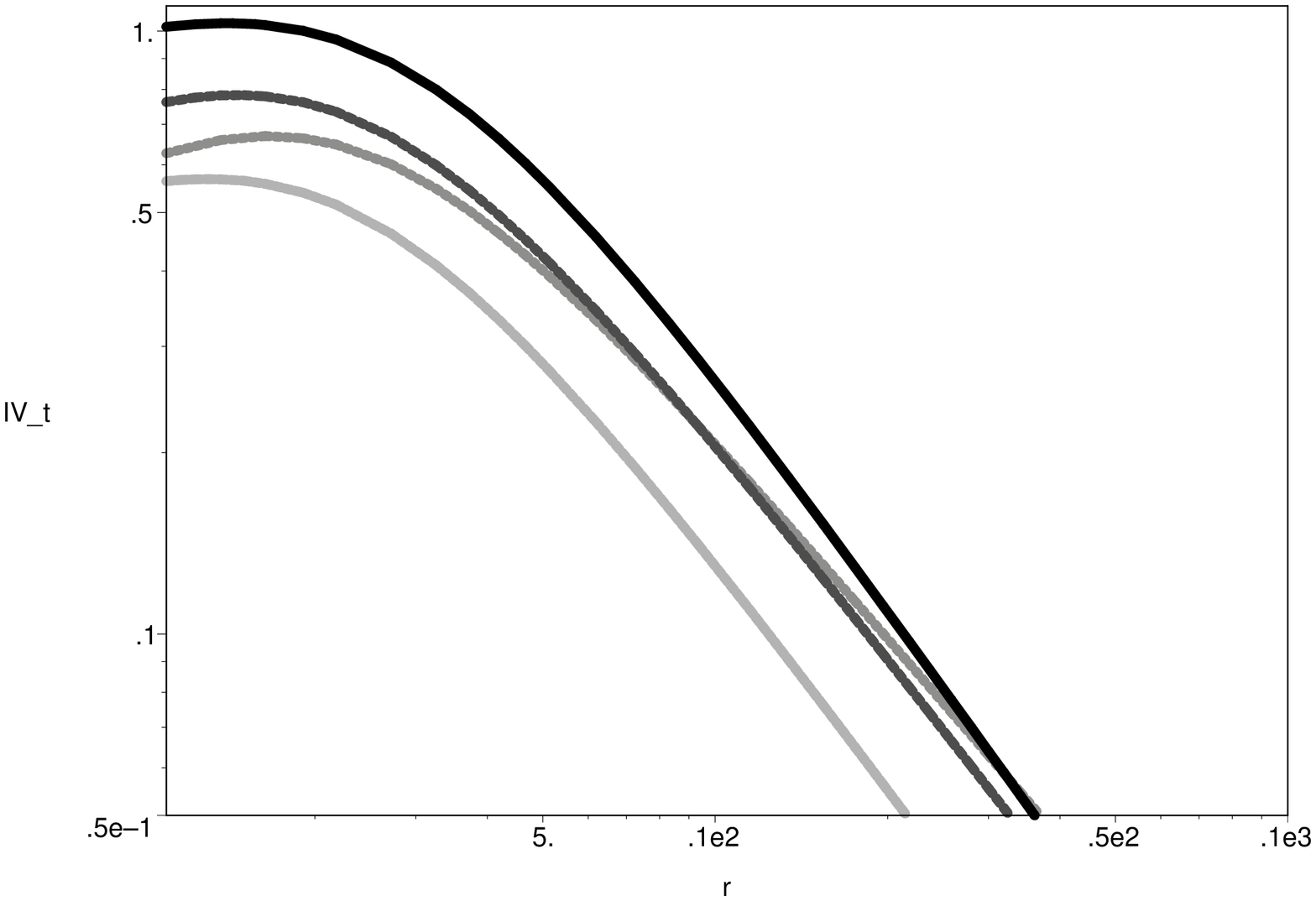}
     \hspace{0.3cm}
    \end{minipage}
\caption{Log-log plots of the isotropic and anisotropic distribution
functions and intrinsic velocity dispersions for the models $\al =
2, \g = 1/3$ (black solid), $\al = 1, \g = 1/2$ (black dashed), $\al
= 2/3, \g = 1$ (light grey), $\al = 1/2, \g = 1/2$ (dark grey);
first row: isotropic and anisotropic distribution function, second
row: isotropic, radial and tangential velocity dispersion for $r
\leq 1$, third row: isotropic, radial and tangential velocity
dispersion for $r > 1$; $G=1=b, r_a=2$}\label{fig}
\end{figure*}

\section{Intrinsic velocity dispersions}

The intrinsic velocity dispersions (VDs) for the isotropic models
are derived from the usual relation
 \bdi
  \langle v_r^2(r)\rangle =
  \f{1}{\rho(r)}\int_r^{\infty}\f{GM(s)\rho(s)ds}{s^2}
   \edi
as follows: Using equation (\ref{ch46}) for the radial range $r>b$
after substituting $s^{\al}=x$, we get
 \beq \label{eq1} \langle v_r^2(r)
\rangle = \Phi_0\g b^{\al
\g}\f{(b^{\al}+r^{\al})^{\g+2}}{[(1+\al)b^{\al}+(1-\al \g)r^{\al}]}
 \left\{ \f{(1+\al)b^{\al}}{1+\f{2}{\al}+2\g}r^{-2\al(1+\g)} \cdot ~
 _2F_1
\left(2\g+3,1+\f{2}{\al}+2\g;2+\f{2}{\al}+2\g;-\left(\f{b}{r}\right)^{\al}\right)
\right.
 \eeq
  \bdi
  \left. + \f{1-\al \g}{\f{2}{\al}+2\g}r^{-\al(2\g+1)} \cdot ~ _2F_1
\left(2\g+3,\f{2}{\al}+2\g;1+\f{2}{\al}+2\g;-\left(\f{b}{r}\right)^{\al}\right)\right\}
\edi whereas for the inner range $r<b$, suitable variable
transformations and usage of equation (\ref{hypergeom2}) results
into
  \beq \label{eq2}
 \langle v_r^2(r) \rangle = \Phi_0\g b^{\al \g}\f{(b^{\al}+r^{\al})^{-\g-1}}{[(1+\al)b^{\al}+(1-\al \g)r^{\al}]}
 \times
 \eeq
 \bdi
\left\{ \f{(1+\al)b^{\al}}{1+\f{2}{\al}+2\g}r^{\al}\cdot ~ _2F_1
\left(2\g+3,1;2+\f{2}{\al}+2\g;\f{b^{\al}}{b^{\al}+r^{\al}}\right) +
\f{1-\al \g}{\f{2}{\al}+2\g}r^{2\al} \cdot ~ _2F_1
\left(2\g+3,1;1+\f{2}{\al}+2\g;\f{b^{\al}}{b^{\al}+r^{\al}}\right)\right\}.
\edi
 In contrast to the DFs from above, the expressions here
involve only special hypergeometric functions of the form
$_2F_1(a,b;c;z).$
 In Fig. 1, second and third row, left plots, we show IV := $\sqrt{\langle v_r^2(r)\rangle}$ for $r \leq 1$ and $r>1$
 respectively for the same model parameters as above. For $r \to 0$,
 the velocity dispersions even
 decrease for the weak cuspy models we are studying (this sounds
counterintuitive, but that is a typical
 behaviour of such models as long as no additional central black hole potential is
 added, see e.g. \cite{trem}): For $r \to 0$, the VDs converge to zero as $r^{2\al}$ if $\al < 2$. For
 $\al = 2$, they are asymptotically constant, as expected: Note that
 the first hypergeometric function in (\ref{eq2}) then simplifies to
 $(b^2+r^2)/r^2$ and the second one to
 $(b/r)^2(1+(b/r)^2)(1+1/(2(\g+1)))$, and both denominators cancel with
 the factor in front.
 Moreover, the outer falloff also
 depends on the degree of the central cusp: The overall shape is flatter for
 increasing cuspiness, but this is already evident from the expression
 for the circular velocity given in (\ref{kreis}).
It can be shown that the projected velocity dispersions exhibit the
same overall behaviour with regard to the model parameters as do the
intrinsic ones.
\\
Now we turn to the Osipkov-Merritt models: The intrinsic
\textit{radial} velocity dispersion is given by (see \cite{carollo})
  \bdi
\langle v_r^2(r)\rangle_a = \f{1}{\rho(r)}\f{r_a^2}{r_a^2+r^2}
  \left\{ \int_r^{\infty}\f{GM(s)\rho(s)}{s^2}ds +
\f{1}{r_a^2}\int_r^{\infty}GM(s)\rho(s)ds \right\} \edi
 where
the first integral was already evaluated in equations (\ref{eq1})
and (\ref{eq2}). The second integral, however, can be evaluated for
$r > b$ as
 \bdi
 \f{1}{r_a^2}\int_r^{\infty}GM(s)\rho(s)ds = \f{\Phi_0 \g b^{\al
 \g}}{r_a^2}\f{b^{\al \g + 2}}{1+\al}
  \times
  \edi
  \bdi
  \left\{
\f{(1+\al)b^{\al}}{1+2\g}r^{-\al(1+2\g)}\cdot ~ _2F_1
\left(2\g+3,1+2\g;2+2\g;-\left(\f{b}{r}\right)^{\al}\right) +
\f{1-\al \g}{2\g}r^{-2\al \g}\cdot ~ _2F_1
\left(2\g+3,2\g;1+2\g;-\left(\f{b}{r}\right)^{\al}\right)\right\}
\edi and for $r<b$ as
 \bdi \f{1}{r_a^2}\int_r^{\infty}GM(s)\rho(s)ds = \f{\Phi_0 \g b^{\al
 \g}}{r_a^2}\f{1}{(b^{\al}+r^{\al})^{2\g+3}}\f{b^{\al \g +2}}{1+\al}
 \times
 \edi
 \bdi
  \left\{
\f{(1+\al)b^{\al}}{1+2\g}r^{2\al}\cdot ~ _2F_1
\left(2\g+3,1;2+2\g;\f{b^{\al}}{b^{\al}+r^{\al}}\right) + \f{1-\al
\g}{2\g}r^{3\al}\cdot ~ _2F_1
\left(2\g+3,1;1+2\g;\f{b^{\al}}{b^{\al}+r^{\al}}\right)\right\}
 \edi
  by using formulae (\ref{ch46}) and (\ref{hypergeom2}) after suitable substitutions, respectively.
 The intrinsic \textit{tangential} velocity dispersion, on the other hand, is then simply given by
 \bdi \langle
v_{\perp}^2(r)\rangle = \f{2r_a^2}{r_a^2+r^2}\langle v_r^2(r)
\rangle_a \edi
 for the respective radial range.
  In Fig. 1, second and third row, central and right plots, we show the radial and tangential velocity dispersion IV$_r$ := $\sqrt{\langle v_r^2(r)\rangle_a}$
  and IV$_t$ := $\sqrt{\langle v_{\perp}^2(r)\rangle}$ respectively, for the same parameters as for the isotropic models.
Both dispersions decrease more slowly for small values of $\gamma$,
i.e. for higher anisotropies. Concerning the overall shape, it can
be shown that IV$_t$ falls off more rapidly for $r>b$ than IV$_r$.
For increasing cuspiness, the shape of both velocity dispersions
becomes flatter, although this behaviour is much less pronounced
than it is for the isotropic models. The tangential velocity
dispersion
 dominates over the radial velocity dispersion for $r_a > b$, whereas the opposite is true for $r_{a} \leq b$.
Since $\langle v_{\perp}^2(r)\rangle \to 2\langle v_r^2(r)
\rangle_a$ for $r \to 0$, the respective curves differ almost only
by an overall factor of two for small radii.

\section{Models with central black hole}

In the presence of a central black hole of point mass $M_{BH}$, the
density $\rho(r)$ is not changed but the potential is modified
according to
  \beq \label{bh}
  \Psi^{\bullet}(r) = \Psi(r) + \f{G M_{BH}}{\Phi_0}\f{1}{r} =: \Psi(r) + \f{\mu}{r}.
   \eeq
The distribution function for the model (\ref{inipot}),
(\ref{density}), (\ref{bh}) can be determined analytically only if
the energy $\mc{E}$ is large, i.e. close to the black hole. Since
$\Psi^{\bullet}(r)$ is no longer invertible with respect to $r$, one
may perform the following variable transformation in $I(\mc{E})$ of
equation (\ref{eddi}) (see e.g. \cite{trem}):
 \beq \label{itrans}
 I(\mc{E}) = \int_0^{\mc{E}}
 \f{d\rho(\Psi^{\bullet})}{d\Psi^{\bullet}}
 \f{d\Psi^{\bullet}}{\sqrt{\mc{E}-\Psi^{\bullet}}} =
 \int_0^{u(\mc{E})} \f{d\rho(u)}{du}
 \f{du}{\sqrt{\mc{E}-\Psi^{\bullet}(u)}},
 \eeq
where $u = 1/r$ and $u(\mc{E})$ is defined implicitly by
$\Psi^{\bullet}(u(\mc{E})) = \mc{E}$. For large $\mc{E}$ and small
$r$ (i.e. large $u$), we may approximate $\Psi^{\bullet} \to \mu u$
and
 \bdi
  \f{d\rho}{du} \to (2-\al)b(u b)^{1-\al} + \f{2(1-\al
  \g)(1-\al)}{1+\al}b(u b)^{1-2\al}.
   \edi
Inserting this into (\ref{itrans}) and using (\ref{ch47}), the
distribution function becomes then
  \beqn
\lefteqn{f^{\bullet}(\mc{E}) =
\f{1}{\sqrt{8}\pi^2}\left[(2-\al)\left(\f{3}{2}-\al
\right)\left(\f{b}{\mu}\right)^{2-\al}B\left(\frac{1}{2},2-\al\right)\mc{E}^{\f{1}{2}-\al}
+ \right. {}}
 \nonumber\\
 & &{} \left. + 2(1-\al
\g)\left(\f{1-\al}{1+\al}\right)\left(\f{3}{2}-2\al\right)\left(\f{b}{\mu}\right)^{2-2\al}B\left(\frac{1}{2},2-2\al\right)
\mc{E}^{\f{1}{2}-2\al}\right] \eeqn
 which is valid for $\al < 1$, but both terms are non-vanishing only for $\al < 3/4$. However, both
  restrictions favour a cusp of $2-\al > 5/4$ which is steeper as the
 $r^{-1/2}$-cusp expected to be produced by the adiabatic growth of a black hole in the context of
isotropic models. The distribution function
 is hardly affected by $\gamma$ and the system is populated with
 more stars in the very centre, i.e. $f^{\bullet}(\mc{E})$ is larger, with
 decreasing black hole mass $\mu$, as expected.
\\
In order to deduce the isotropic velocity dispersion in the presence
of a black hole, we use the relation
 \bdi \langle
v_r^2(r)\rangle^{\bullet} = \langle v_r^2(r)\rangle +
\f{\mu}{\rho(r)}\int_r^{\infty}\f{\rho(s)}{s^2}ds, \edi
 where
the first term is the velocity dispersion from equ. (\ref{eq1}) and
(\ref{eq2}), and the second term can be evaluated as
 \bdi
 \f{\mu}{\rho(r)}\int_r^{\infty}\f{\rho(s)}{s^2}ds = \f{\mu}{\al[(1+\al)b^{\al}+(1-\al \g)r^{\al}]}
 \left\{ \f{(1+\al)b^{\al}}{1+\f{3}{\al}+\g}r^{-1}\cdot ~ _2F_1
\left(\g+2,1;2+\f{3}{\al}+\g;\f{b^{\al}}{b^{\al}+r^{\al}}\right) +
\right.
 \edi
  \bdi
 \left.
\f{1-\al \g}{\f{3}{\al}+\g}r^{\al-1}\cdot ~ _2F_1
\left(\g+2,1;1+\f{3}{\al}+\g;\f{b^{\al}}{b^{\al}+r^{\al}}\right)\right\}
 \edi
 for $r<b$, and
\bdi
 \f{\mu}{\rho(r)}\int_r^{\infty}\f{\rho(s)}{s^2}ds = \f{\mu (b^{\al}+r^{\al})^{\g +2}}{\al[(1+\al)b^{\al}+(1-\al \g)r^{\al}]}
  \left\{ \f{(1+\al)b^{\al}}{1+\f{3}{\al}+\g}r^{-1-\al(2+\g)}\cdot ~
 _2F_1
\left(\g+2,1+\f{3}{\al}+\g;2+\f{3}{\al}+\g;-\left(\f{b}{r}\right)^{\al}\right)
\right.\edi \bdi
 \left. + \f{1-\al \g}{\f{3}{\al}+\g}r^{-1-\al(1+ \g)} \cdot ~ _2F_1
\left(\g+2,\f{3}{\al}+\g;1+\f{3}{\al}+\g;-\left(\f{b}{r}\right)^{\al}\right)\right\}
 \edi
 for $r > b$ by using (\ref{ch46}) and (\ref{hypergeom2}), respectively.
 In Fig. 2, we show IV$_b := \sqrt{\langle v_r^2(r)\rangle^{\bullet}}$
  for $r \leq 1$ and $r > 1$. Now the velocity dispersions
rise steeply for $r \to 0$ in contrast to the previous case without
black hole. It can be also shown that this rise at small radii is
more pronounced if $\mu$ is increased but $\al$ is fixed
\textit{together} with a more slowly falloff in the outer parts. The
same overall behaviour is found for the corresponding projected
velocity dispersions.
\begin{figure*}[hb]
\begin{minipage}[b]{0.4\linewidth}
      \centering\includegraphics[angle=270,width=5cm]{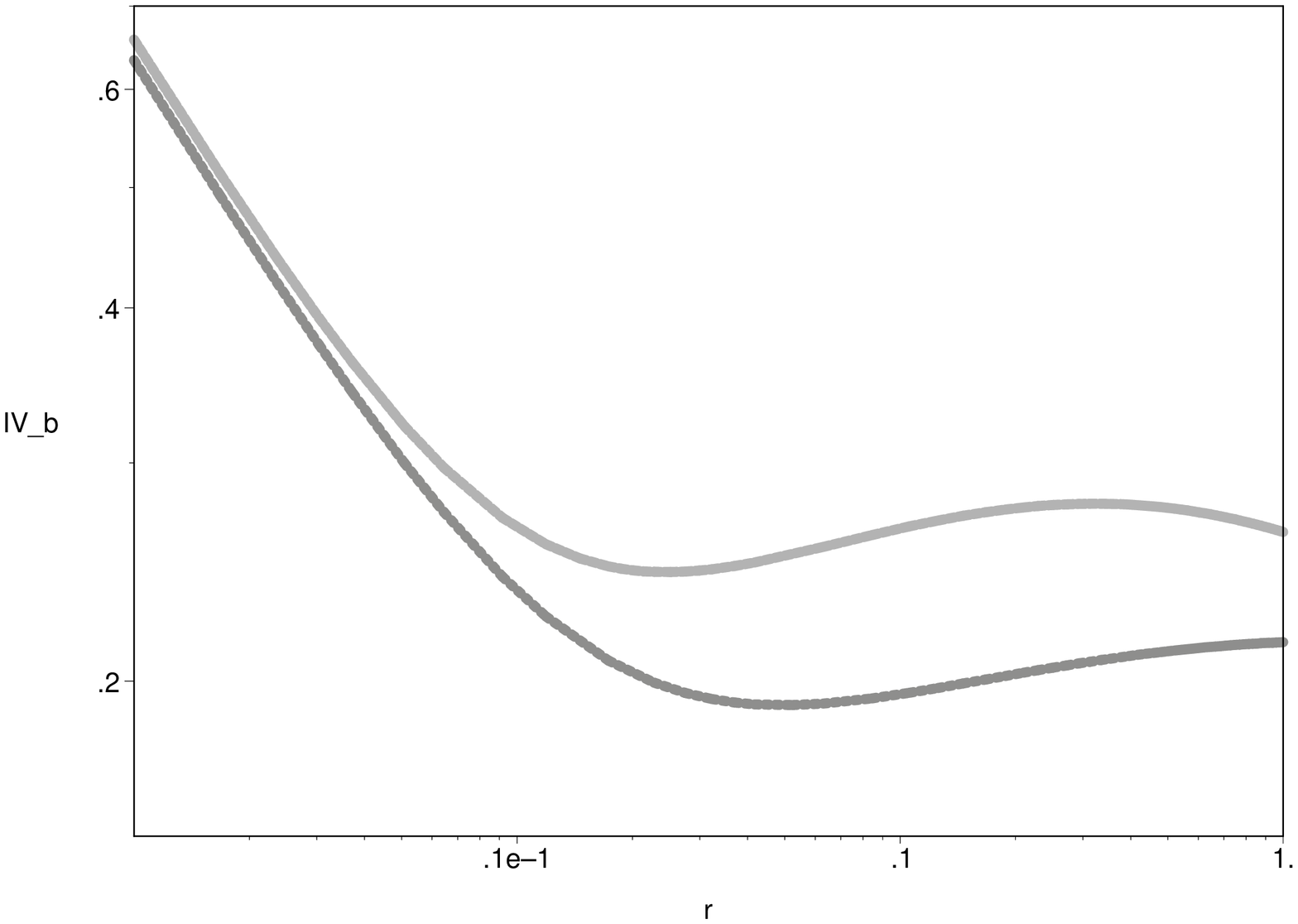}
     \hspace{0.1cm}
    \end{minipage}%
     \begin{minipage}[b]{0.4\linewidth}
      \centering\includegraphics[angle=270,width=5cm]{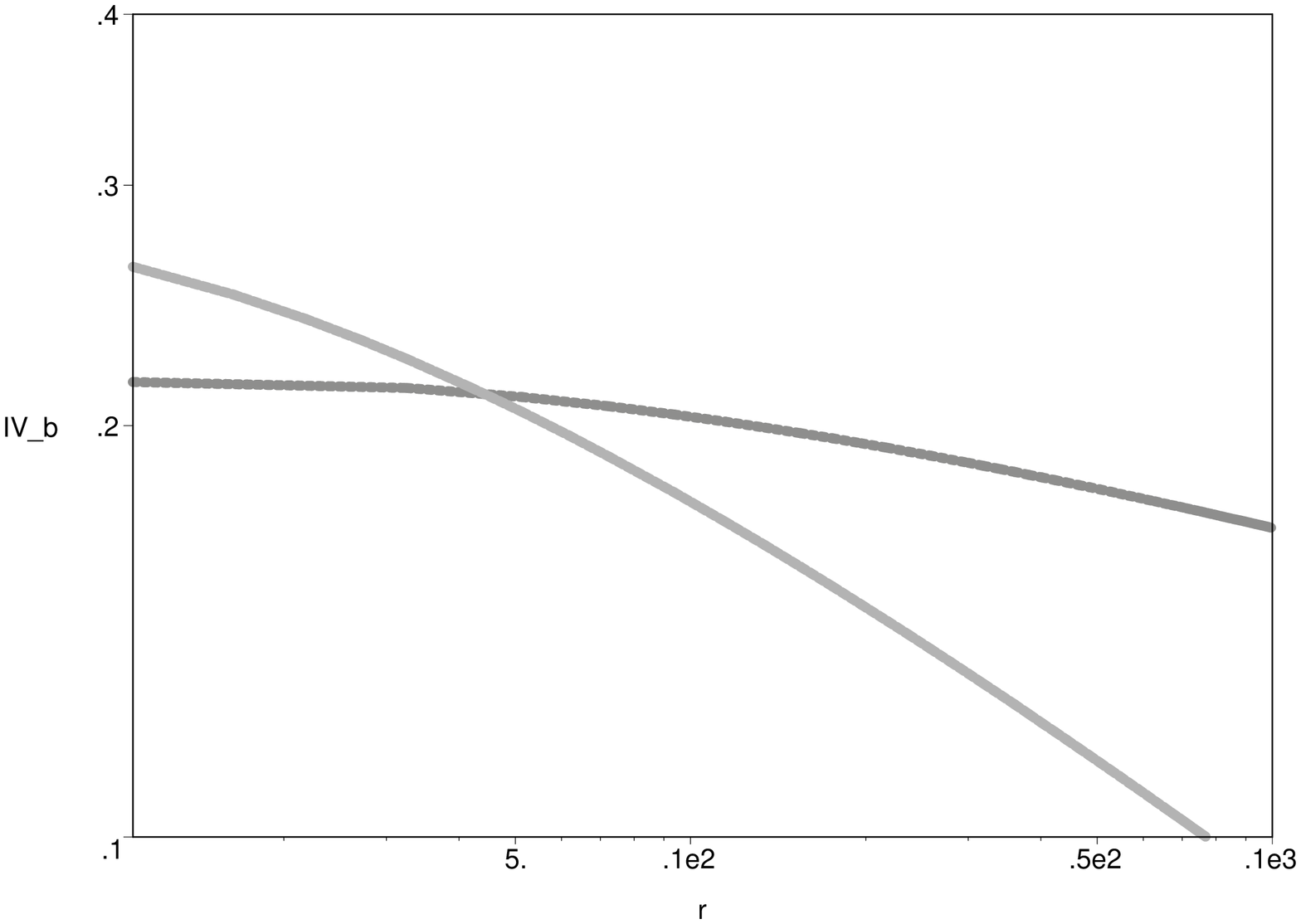}
     \hspace{0.1cm}
    \end{minipage}
\caption{Log-log plots of the intrinsic velocity dispersion for
models with black hole: $\al = 2/3, \g = 1$ (light grey), $\al =
1/2, \g = 1/2$ (dark grey), $\mu = 0.001$, $G=1=b$}\label{fig2}
\end{figure*}

\section{Conclusions}

In this paper, we considered a family of non-singular potentials
falling off as $1/r$ or more slowly at large radii. The associated
self-consistent mass density incorporates flat or cuspy nuclear
regions together with a flexible falloff behaviour at large
distances from the centre. The corresponding distribution functions
and intrinsic velocity dispersions can be represented analytically
in terms of hypergeometric functions. This allows a straightforward
comparison between models for galaxies having different central and
outer shapes in the mass density. We restricted ourselves to
isotropic and anisotropic models of Osipkov-Merritt type. It is
shown that the anisotropy affects the distribution functions only
outside the central parts where they do not fall off as rapidly as
the isotropic ones, whereas the increase for large arguments is
dominated by the cusp parameter in both cases. Moreover, the
velocity dispersions decrease more rapidly for the less anisotropic
models and their shape is flatter for increasing cuspiness. The
presence of a central point mass potential, mimicking a massive
black hole, is also studied. It is shown that the velocity
dispersions rise steeply at small radii for increasing black hole
mass, which in the same time leads to higher values of the velocity
dispersion over a wider radial range.

\appendix

\section{Formulae}

The formula for the derivative of the general hypergeometric series
is applied in \textit{Section 3},
 \beq \label{hypergeom}
  \f{d}{dz} ~_pF_q (a_1,a_2,...,a_p;b_1,b_2,...,b_q;z) = \f{a_1 a_2 ... a_p}{b_1 b_2 ...
  b_q}~ _pF_q (a_1+1,a_2+1,...,a_p+1;b_1+1,b_2+1,...,b_q+1;z).
 \eeq
In \textit{Section 4}, the transformation formula for the special
hypergeometric function is used
 \beq \label{hypergeom2}
 _2F_1 (a,b;c;z) = (1-z)^{-a}~ _2F_1 \left(a,c-b;c;\f{z}{z-1}\right).
 \eeq
 Following integral relations are
used in the text, see \cite{grad}:
\begin{eqnarray} \label{ch47}
\lefteqn{\int_{0}^{u}x^{\nu-1}(u - x)^{\mu-1}(x^{m} +
\beta^{m})^{\lambda}dx = \beta^{m\lambda}u^{\mu+\nu-1}B(\mu,\nu)
\times {}}
\nonumber\\
& &{} ~_{m+1}F_{m}\left(-\lambda,\frac{\nu}{m},\frac{\nu+1}{m},...,
\frac{\nu+m-1}{m};\frac{\mu+\nu}{m},\frac{\mu+\nu+1}{m},...,\frac{\mu+\nu+m-1}{m};-\left(\frac{u}{\beta}\right)^{m}\right)
\end{eqnarray}
if $\rm{Re}(\mu) > 0,~ \rm{Re}(\nu) > 0$.
\begin{equation}  \label{ch46}
\int_{u}^{\infty}x^{-\lambda} (x + \beta)^{\nu} (x - u)^{\mu-1}dx =
u^{\mu+\nu-\lambda}B(\lambda-\mu-\nu,\mu)~_{2}F_{1}
\left(-\nu,\lambda-\mu-\nu;\lambda-\nu;-\frac{\beta}{u}\right)
\end{equation}
if $0 < \rm{Re} (\mu) < \rm{Re} (\lambda-\nu)$.

\label{lastpage}

\end{document}